\documentclass[reprint, aps, pra, 
nofootinbib]{revtex4-2}

\usepackage{tabularx}
\usepackage{xcolor} 
\usepackage[T1]{fontenc}

\newcounter{Protocol}

\usepackage{amsthm}

\newtheorem*{Game*}{Game}
\newtheorem*{TokGen*}{TokenGeneration Phase}
\newtheorem*{TokVer*}{TokenVerification Phase}
\newtheorem*{Setup*}{Setup}
\newtheorem*{Query*}{Query}
\newtheorem*{Challenge*}{Challenge}
\newtheorem*{Guess*}{Guess}
\usepackage{breqn}
\theoremstyle{remark}

\usepackage{algorithm}
\usepackage{algpseudocode}

\usepackage{amssymb}

\usepackage{comment}

\usepackage{tikz}
\usetikzlibrary{quantikz}

\usepackage[hidelinks]{hyperref}

\usepackage[capitalize]{cleveref}
\usepackage{tensor}
\usepackage{algpseudocode}
\usepackage{mathtools, bbm}
\usepackage{graphicx}
\usepackage{dcolumn}
\usepackage{bm}

\usepackage{chemformula} 
\setchemformula{
    format = \rmfamily,
}

\usepackage{siunitx} 
\DeclareSIUnit{\belmilliwatt}{Bm}

\usepackage[shortcuts,acronym]{glossaries} 
\setkeys{glslink}{hyper=false} 

\newacronym{bs}{BS}{Beam Splitter}
\newacronym{mzi}{MZI}{Mach-Zehnder Interferometer}
\newacronym{ps}{PS}{Phase Shifter}
\newacronym{tbs}{TBS}{Tunable Beam Splitter}
\newacronym{adc}{ADC}{Analog Digital Converter}
\newacronym{nisq}{NISQ}{Noisy Intermediate-Scale Quantum}
\newacronym{qaoa}{QAOA}{Quantum Approximate Optimization Algorithm}
\newacronym{vqe}{VQE}{Variational Quantum Eigensolver}
\newacronym{vqc}{VQC}{Variational Quantum Circuit}

\newcommand{\eg}{\textit{e.g.}~} 
\newcommand{\ie}{\textit{i.e.}~} 

\definecolor{TUMAccentOrange}{HTML}{E37222}

\begin{document}

\preprint{APS/123-QED}

\title{Photonic processor benchmarking for variational quantum process tomography}

\author{Vladlen Galetsky$^{*}$, Paul Kohl$^{*}$, Janis N{\"o}tzel$^{*}$}%
\affiliation{%
 $^{*}$Emmy-Noether Group Theoretical Quantum Systems Design$,$ Technical University of Munich$,$ 80333 Munich$,$ Germany%
}

\date{\today}

\begin{abstract}
We present a quantum-analogous experimental demonstration of variational quantum process tomography using an optical processor. This approach leverages classical one-hot encoding and unitary decomposition to perform the variational quantum algorithm on a photonic platform. We create the first benchmark for variational quantum process tomography evaluating the performance of the quantum-analogous experiment on the optical processor against several publicly accessible quantum computing platforms, including IBM’s 127-qubit Sherbrooke processor, QuTech’s 5-qubit Tuna-5 processor, and Quandela’s 12-mode Ascella quantum optical processor. We evaluate each method using process fidelity, cost function convergence, and processing time per iteration for variational quantum circuit depths of 
$d=3$ and
$d=6$. Our results indicate that the optical processors outperform their superconducting counterparts in terms of fidelity and convergence behavior reaching fidelities of $0.8$ after $9$ iterations, particularly at higher depths, where the noise of decoherence and dephasing affect the superconducting processors significantly. 

We further investigate the influence of any additional quantum optical effects in our platform relative to the classical one-hot encoding. From the process fidelity results it shows that the (classical) thermal noise in the phase-shifters dominates over other optical imperfections, such as mode mismatch and dark counts from single-photon sources.

The benchmarking framework and experimental results demonstrate that photonic processors are strong contenders for near-term quantum algorithm deployment, particularly in hybrid variational contexts. This analysis is valuable not only for state and process tomography but also for a wide range of applications involving variational quantum circuit based algorithms. 
\end{abstract}

\maketitle

\section{Introduction}
\label{Introduction}
Classical and quantum optical processors offer several advantages over alternative quantum hardware platforms, including ion traps~\cite{Liu2025}, superconducting qubits~\cite{Arute2019}, and Rydberg atom arrays~\cite{Bluvstein2024}. One key advantage is their compatibility with both quantum and classical photonic infrastructures, especially those operating at telecom wavelengths (\eg the C-Band around \qty{1550}{\nano\meter}), making them promising candidates for distributed quantum computing~\cite{Taballione_2021}. While ion- and atom-based systems typically require cryogenic cooling at temperatures below \qty{20}{\milli\kelvin}, photonic processors often can operate at room temperature. Moreover, since photons interact only weakly with their environment, photonic systems exhibit longer coherence times than superconducting systems, which typically suffer from decoherence and dephasing times on the order of \qty{150}{\micro\second} \cite{mckay2023benchmarkingquantumprocessorperformance}, often necessitating quantum error correction for complex algorithms.

Photonic processors currently achieve high fidelities for Haar-random unitary implementations, reaching \qty{97.4}{\percent} and \qty{98.6}{\percent} in 12-mode and 20-mode devices, respectively, with corresponding photon losses per mode of approximately 
\qtyrange{43.8}{54.3}{\percent}.~\cite{Taballione_2021,degoede2022highfidelity12modequantum}
Recent developments in integrated photonics include the creation of the first Gottesman-Kitaev-Preskill (GKP) qubit by Xanadu~\cite{Larsen2025} and a demonstration of quantum computational advantage via Gaussian boson sampling on the Borealis processor~\cite{Madsen2022}. New applications continue to emerge, such as the proposal to simulate real-time dynamics of quantum field theories using an optical, measurement-based continuous-variable processor~\cite{PhysRevA.110.012607}.

A particularly promising application is in variational quantum computation, where hybrid quantum-classical approaches leverage parametrized quantum circuits (\ie \glspl{vqc}) on \gls{nisq} devices to perform optimization tasks in a hardware-in-the-loop manner. \glspl{vqc} form the backbone of quantum machine learning frameworks, including quantum deep neural networks and kernel-based methods~\cite{Beer2020,Jerbi2023}. They also play a central role in algorithms such as the \gls{qaoa}~\cite{farhi2014quantumapproximateoptimizationalgorithm}, the \gls{vqe}~\cite{Tilly_2022}, and variational approaches for simulating many-body Hamiltonians in condensed matter physics~\cite{landman2022classicallyapproximatingvariationalquantum}.

In the context of quantum state and process tomography, \glspl{vqc} provide resource-efficient alternatives to traditional measurement-based methods~\cite{PhysRevA.77.032322}. \citeauthor{PhysRevA.87.032304}~first proposed variational quantum process tomography using an optical quantum processor \cite{PhysRevA.87.032304}, approximating the process matrix $\chi$ variationally to fit experimental data. Their method matched the accuracy of standard techniques while reducing the required number of measurements from 256 to just 55. In another application, \citeauthor{Carolan_2020}~\cite{Carolan_2020} used a quantum optical processor to variationally compile and invert an unknown unitary operation using a limited number of input states. More recently, \citeauthor{PhysRevLett.129.133601}~demonstrated an entanglement-assisted approach to process tomography on an optical processor \cite{PhysRevLett.129.133601}, reducing the number of required input states from $4^n$ to $2^{n-m}$ however requiring entanglement based state initialization as well as using $m$ auxiliary qubits.

\section{Our contributions}
\label{Our contributions}

In this work, we propose utilizing the architecture introduced in~\cite{Galetsky_2024}, adapted to a local optical processor employing one-hot encoding to emulate the behavior of a quantum optical processor, as illustrated in Fig.~\ref{UV_Fig}. This approach reduces the number of required input state initializations from the standard $4^n$ to $2^n$ orthogonal states per parameter update. For an $n$-qubit system, this one-hot encoding scheme requires a total of $2^{t+1}$ optical modes, with $t$ being the target qubit dimension.

Due to the limited benchmarking available across different quantum hardware platforms, this study aims to be the first to evaluate the applicability of variational quantum process tomography across various \gls{nisq} devices. Specifically, we analyze cost function behavior, process fidelity, and estimated processing time for two different circuit depths applied to Haar-random two-qubit unitary processes. We consider $d=6$ as the optimal depth for this qubit size, following the definition in~\cite{Galetsky_2024}. The platforms evaluated include our local optical processor, the Tuna 5-qubit superconducting processor from QuTech~\cite{valléssanclemente2025optimizingfrequencypositioningtunable}, IBM's 127-qubit Sherbrooke processor~\cite{mckay2023benchmarkingquantumprocessorperformance}, and the Ascella quantum optical processor developed by Quandela~\cite{Maring2024}. This comparison is not only relevant for process tomography but also provides insights into the general utility of other \gls{vqc} based algorithms such as \gls{qaoa} and \gls{vqe} in \gls{nisq} devices.

Furthermore, we investigate how effectively one-hot encoding implemented on a classical optical processor can approximate the behavior of a genuine quantum optical processor. Despite the presence of hidden quantum noise sources such as dark counts in single-photon detectors and mode mismatch, from the comparison of the cost function and process fidelity our analysis indicates that these effects are negligible as the noise effects are comparable between the two devices ~\cite{Rohde_2006}.

Our experimental results show that, for both quantum and classical optical processors, increasing the circuit depth from $d=3$ to $d=6$ improves process fidelity from \qty{0.71}{} to \qty{0.80}{}, while reducing the cost function from \qty{0.35}{} to \qty{0.22}{}. In contrast, the performance of superconducting processors did not improve beyond four iterations due to cumulative decoherence and dephasing effects introduced by the increased circuit depth. Additionally, the initial average process fidelity for these platforms degraded from \qty{0.42}{} to \qty{0.30}{} as depth increased, likely due to a larger parameter space of the process matrix being represented by deeper \glspl{vqc}.

Finally, we note that the average processing time per iteration across all platforms for \( d = 3 \) is approximately \qty{1400}{\second}. However, there are notable outliers: the local optical processor and the 5-qubit \textit{Tuna} processor, which exhibit significantly shorter and longer processing times of up to \qty{400}{\second} and \qty{4400}{\second}, respectively. The shorter processing times observed for the local optical processor may be attributed to the absence of required sampling, as its intensity measurements directly reflect the amplitude distribution of the results.

\section{Notations}
\label{Notations}
The notations and theoretical framework for the algorithm follow the work presented in \cite{Galetsky_2024}.
The authors adopt a vector notation, such as \( \hat{\theta} \), where \( \hat{\theta}_i \) denotes the \( i \)-th block of the overall parameter vector \( \hat{\theta} \). Similarly, \( \theta_i \) refers to the \( i \)-th individual parameter \( \theta \). Quantum gates are represented by uppercase letters, \eg RY(\( \theta \)), and sequences of quantum gates are connected using hyphens, for example: RY(\( \theta_0 \))-RX(\( \theta_1 \)).

Pure quantum states are expressed using Dirac bra-ket notation, \( \ket{\psi} \), while general density matrix states are denoted using lowercase Greek letters, such as \( \rho \) and \( \sigma \). The number of qubits is indicated by \( n \), with \( t \) and \( a \) denoting the number of target and ancilla qubits, respectively.

\section{Methods}
\label{Methods}

This work follows the extension of the work in \cite{Galetsky_2024} but now in quantum hardware. To summarize, we introduce a process tomography algorithm with architecture shown in Fig.\,\ref{architecture}.

The \gls{vqc} algorithm is composed of single qubit sequences RZ($\theta_{i}$)-RY($\theta_{i+1}$)-RZ($\theta_{i+2}$) with a CX gate alternating between odd or even qubits depending on the index of the repetition block.

The cost function used to optimize \( U_{\text{VQC}}(\hat\theta) \) is defined as:
\begin{align}
C(\hat{\theta}) &= \frac{1}{2^n} \sum_{i=1}^{2^n} \| \ket{\Psi_i^{\mathrm{pr}}} - \ket{\Psi_i^{\mathrm{tr}}} \|^2  \notag \\
&= \frac{1}{2^{n-1}} \sum_{i=1}^{2^n} \left[1 - \Re \left( \braket{\Psi_i^{\mathrm{tr}}}{\Psi_i^{\mathrm{pr}}} \right) \right]. 
\label{costfun}
\end{align}
Where $\ket{\Psi_{i}^{\mathrm{pr}}} = U\textsubscript{VQC}(\hat{\theta})\ket{i}$
and $\ket{\Psi_{i}^{\mathrm{tr}}} = U\ket{i}$, with the four-term shift rule being used to optimize the $\hat\theta$ parameters in between of each iteration of the \gls{vqc} architecture:

\begin{align}
\nabla_{\theta_i} C(\hat{\theta}) &=  
\frac{\sqrt{2}+1}{4\sqrt{2}} \left[ C(\hat{\theta}) \Big|_{\theta_i + \frac{\pi}{2}} - C(\hat{\theta}) \Big|_{\theta_i - \frac{\pi}{2}} \right] \notag\\
&\quad - \frac{\sqrt{2}-1}{4\sqrt{2}} \left[ C(\hat{\theta}) \Big|_{\theta_i + \frac{3\pi}{2}} - C(\hat{\theta}) \Big|_{\theta_i - \frac{3\pi}{2}} \right]. 
\label{parameter}
\end{align}

The target unitary \( U \) is generated from the Haar measurement using the QR decomposition method \cite{roberts2020qrlqdecompositionmatrix}. Optimization is performed using the Adam optimizer, with hyperparameters \( \beta_1 = 0.8 \) and \( \beta_2 = 0.999 \), representing the decay rates for the first and second moment estimates, respectively.

To compare the performance of different quantum devices, we use the process fidelity metric. In this context, every completely positive (CP) quantum operation is associated with a process matrix \( \chi \), which represents the operation in the Chi (or Choi) representation. Our analysis is based on this representation, and the process fidelity is defined as:

\begin{equation}
  F_{\text{process}} = \mathrm{Tr}\left( \chi_{\mathcal{E}_{\text{actual}}} \chi_{\mathcal{E}_{\text{ideal}}} \right)
  \label{processfid}
\end{equation}

where \( \chi_{\mathcal{E}_{\text{actual}}} \) and \( \chi_{\mathcal{E}_{\text{ideal}}} \) are the process matrices corresponding to the actual and ideal quantum operations, respectively.

\section{Choice of quantum hardware}
\label{Choice of quantum hardware}

To evaluate the performance of the local optical processor and assess the effectiveness of the classical one-hot encoding within the photonic framework, we benchmarked it against the Ascella 12-mode quantum optical processor developed by Quandela, which features a comparable architecture.

Ideally, a comprehensive benchmarking would include comparisons to a broader set of platforms such as ion or atomic traps, Rydberg atom systems, and superconducting qubit processors. However, due to the limited availability of open-access hardware, no ion/atomic trap or Rydberg-based quantum devices were accessible at the time of this study.

Given these constraints, we selected IBM’s 127-qubit Sherbrooke processor, and Tuna-5, a 5-qubit superconducting processor developed by QuTech. Although both Sherbrooke and Tuna-5 are superconducting platforms, they differ significantly in architecture: IBM employs a square-lattice topology, while Tuna-5 utilizes a star-shaped coupling map. This structural difference impacts gate transpilation, connectivity, and noise profiles, particularly leakage and crosstalk during two-qubit gate operations such as controlled-X (CX) gates. Appendix\,\ref{Sec:Appendix} provides detailed information on the hardware specifications and coupling maps for each system.

We initially intended to include benchmarking results from a Nuclear Magnetic Resonance (NMR) quantum processor \cite{Jones_2024} developed by SpinQ. However, the hardware was unavailable during the course of this study. Additionally, while the X8 quantum optical processor by Xanadu is freely accessible \cite{Arrazola2021}, its interferometer-based architecture does not support non-Gaussian gates, thereby lacking a universal gate set. Furthermore, the device is partitioned into two subsystems, effectively restricting operations to the form \( \mathrm{SU}(4) \otimes \mathrm{SU}(4) \), which limits the applicability of our algorithm on this platform.

\section{Optical processor hardware}
\label{Optical processor hardware}

\subsection{Photonic Processor Working Principle}

The local optical processor is essentially a multiport interferometer based on the scheme by \citeauthor{ClementsUniversalMultiportInterferometers} \cite{ClementsUniversalMultiportInterferometers}, that means it is a mesh of unit cells, each built from one \gls{tbs} and one external \gls{ps} on two optical modes that successively build up a bigger unitary scattering matrix on a given number of (spatial) modes. The local optical processor used in this study has 8 modes, see Fig.~\ref{UV_Fig}.
It is integrated on a single chip consisting of silicon nitride (\ch{Si3N4}) waveguides. The \gls{tbs} are realised as \glspl{mzi}, \ie they consist of a fixed 50:50 \gls{bs} then a \gls{ps} on one of the output paths and then recombining the two paths of equal length with another fixed 50:50 \gls{bs}. This way one can steer the amount of light passing through the \gls{mzi} to each of the outputs by destructive/constructive interference on the second \gls{bs} by varying the phase shift.

\begin{figure}
    \centering
    \begin{quantikz}
        &&\lstick{$\ket{\Psi}$} & \qw & \gate{U} & \qw & \gate{U_{\text{VQC}}} & \qw & \qw \\
        &&\lstick{$\ket{0}$} & \gate{H} & \ctrl{-1} & \qw & \octrl{-1} & \gate{H} & \meter{}
    \end{quantikz}
    \caption{Circuit architecture for the PT\_VQC tomography algorithm described in \cite{Galetsky_2024}.}
    \label{architecture}
\end{figure}
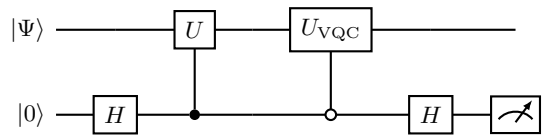

However, the available processor does not implement a full universal multiport interferometer that can implement any unitary transformation on that amount of modes, because one row of \glspl{ps} on the outputs of the processor can not be set (red \glspl{ps} in Fig.~\ref{UV_Fig}). Thus the processor can only implement any unitary up to a phase shift on the outputs when using the total number of available modes (\ie the diagonal matrix of phase shifts $D$ in the scheme by \citeauthor{ClementsUniversalMultiportInterferometers}~\cite{ClementsUniversalMultiportInterferometers} is missing), this configuration is well-suited to the algorithm proposed in \cite{Galetsky_2024} as it evaluates the unknown unitaries up to a global phase. That only means it implements the correct intensity distribution on the output when fed with light at the inputs. 
As we cannot implement a full unitary with different phases on the full chip due to the inactive \glspl{ps}, the processor was supplied with a coherent state produced by an available laser light source. Because the implemented scattering matrix is insensitive to relative phase between the different outputs, the measurement setup also consists of simple intensity measurements, also inherently insensitive to phase.


The angle $\theta$ of the \gls{tbs} of a given unit cell has the following important settings: For total transmission -- \ie swapping the light modes -- one sets $\theta = 0$. For total reflection on the \gls{tbs} one sets $\theta = \pi$. Accordingly, $\theta = \pi/2$ is set to get a 50:50 splitting ratio on the \gls{tbs}. The phases set on the processor are implemented via local heating of the waveguides inducing a change in refractive index in the waveguides. This limits the speed of setting matrices on the processor, because each heater has to reach thermal equilibrium for a stable output. 

The available processor can be controlled by a closed-source library provided by the manufacturer. This enables us to set a unitary on the whole processor directly (up to relative phase differences on the modes). One can also directly set single \glspl{ps} and \glspl{tbs} to angle values to manually route/split the light if necessary.

The scheme by \citeauthor{ClementsUniversalMultiportInterferometers} hinges on the fact, that the unit cell connecting two neighboring modes $m_1$ and $m_2$ performs the following unitary transformation $U_{m_1 m_2}(\theta, \phi) \in \mathrm{U}(m)$ for $m$ total modes:
\begin{equation}
    U_{m_1 m_2}(\theta, \phi) = 
    \begin{bmatrix}
    1 &        &                         &                 &        &   \\
      & \ddots &                         &                 &        &   \\
      &        & e^{i\phi}\cos{\theta/2} & -\sin{\theta/2} &        &   \\ 
      &        & e^{i\phi}\sin{\theta/2} &  \cos{\theta/2} &        &   \\
      &        &                         &                 & \ddots &   \\
      &        &                         &                 &        & 1 
    \end{bmatrix},
\end{equation}
with $\theta \in [0, \pi]$ setting the reflectivity of the \gls{tbs} and $\phi \in [0, 2\pi]$ setting the phase shift between modes.
This assumes the unit cell construction from \cite{ClementsUniversalMultiportInterferometers}. 
In general one can write any unitary $U \in \mathrm{U}(m)$ as
\begin{equation}
    U = D \prod_{(m_1, m_2) \in S} U_{m_1 m_2},
\end{equation}
with $S$ being a specific ordered sequence of the transformations on two modes.
In total, this describes a concatenation of $\frac{m(m-1)}{2}$ unit cells in a grid as shown exemplary for 8 modes in Fig.~\ref{UV_Fig} and $D$ being a complex diagonal matrix with modulus one on the diagonal. The $D$ matrix corresponds to phase shifts on all channels at the output, which cannot be set for the local optical processor \cite{ClementsUniversalMultiportInterferometers}.

The \texttt{set\_unitary()} function from the pre-compiled closed-source control software library provided by the photonic chip manufacturer has to set phase and beam splitting angles analogous to the scheme from \citeauthor{ClementsUniversalMultiportInterferometers} \cite{ClementsUniversalMultiportInterferometers}. Yet, it uses a different arrangement of unit cell in the multiport interferometer mesh, where both \glspl{ps} of a unit cell are acting on the second mode instead of the first and the external \gls{ps} of each cell is acting after the \gls{tbs} instead of before. This changes the decomposition, but has the same capabilities in principle. For repeating our experiment one would follow the instructions given in \cite{ClementsUniversalMultiportInterferometers} to build an equivalent processor with available decomposition method.

\begin{figure*}
\hspace{15mm}
\includegraphics[width=1.0\linewidth]{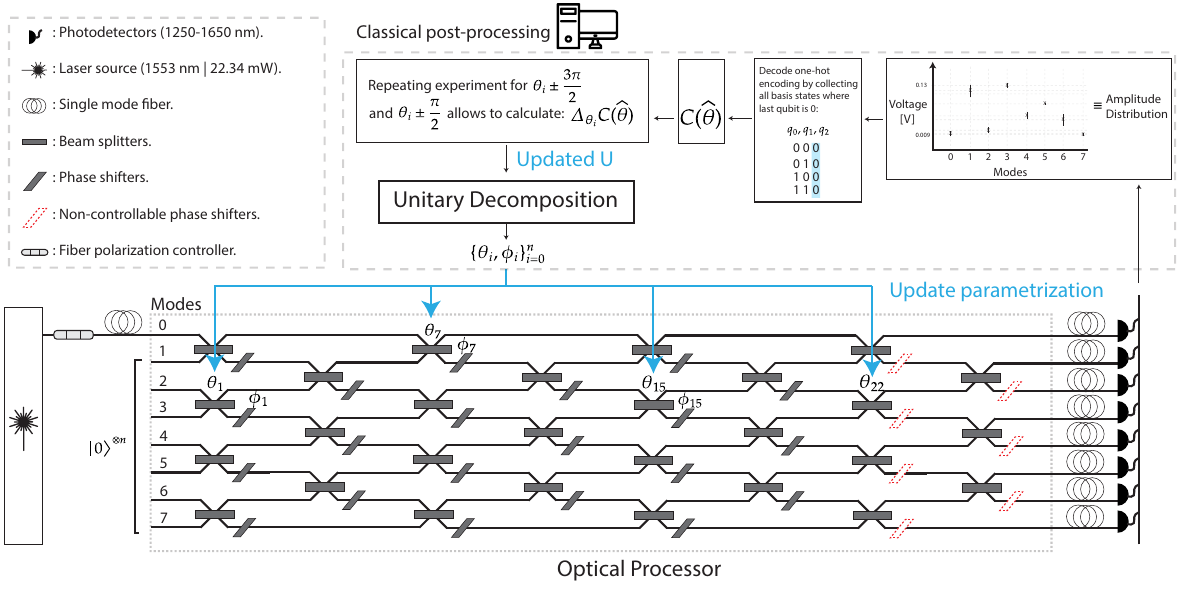}
\caption{We perform a process tomography experiment using an 8-mode optical variational quantum circuit (\gls{vqc}). The procedure begins by converting the quantum circuit shown in Fig.~\ref{architecture} (prior to measurement) into its equivalent unitary representation \( U \). This unitary is then decomposed using a method analogous to the description in~\cite{ClementsUniversalMultiportInterferometers}, yielding the optical parameters \( \{\theta_i, \phi_i\}_{i=0}^{n} \), where \( \theta_i \) and \( \phi_i \) correspond to the beam splitter and phase shifter angles, respectively.
To simulate the experiment with classical hardware, we encode states using a one-hot encoding scheme for which the operation of a quantum optical processor coincides with the classical scattering matrix \cite{LinearOpticsOnlyAllowEveryQuantumOperationForOnePhotonOrOnePath, MethodToDetermineQuantumOperationsRalizableWithLinearOptics}. At the output, photodetectors record voltage signals, which are normalized to obtain the intensity distribution across the modes.
To compute the process tomography cost function \( C(\hat{\theta}) \), we focus on the intensity distribution associated with the last qubit. Specifically, we decode the one-hot encoding by extracting and normalizing the amplitudes of the basis states in which the last qubit is in the \( \ket{0} \) state.
The gradient \( \Delta_{\theta_i} C(\hat{\theta}) \) is estimated using the four-parameter-shift rule described in Eq.\,\ref{parameter}. For each tunable optical parameter \( \{\theta_i, \phi_i\}_{i=0}^{n} \), we evaluate the cost function at shifted values: \( \theta_i \pm \frac{\pi}{2} \), \( \phi_i \pm \frac{\pi}{2} \), \( \theta_i \pm \frac{3\pi}{2} \), and \( \phi_i \pm \frac{3\pi}{2} \), and update accordingly.}
\label{UV_Fig}       
\end{figure*}

\subsection{Information Encoding}

Only a laser as input source was available for experimentation, thus it is crucial to encode the information in a way that does not suffer from phase instabilities when feeding the processor with coherent light in multiple input ports at the same time, which can be a problem \cite{ModelingAnalysisPhaseInstabilityPhotonicProcessor}.
Additionally, the chip itself is only a linear optics device which is limited in its quantum information processing (see \cite{KLM-LOQC, LOQCwithPhotonicQubits, LinearOpticsOnlyAllowEveryQuantumOperationForOnePhotonOrOnePath, MethodToDetermineQuantumOperationsRalizableWithLinearOptics, OptimalApproxToUnitaryQuantumOperatorsWithLinearOptics} for more information on these limitations, on which transformations are possible, and on how to approximate quantum operations with linear optics). Thus the easiest way of utilizing the processor for this purpose is to use so-called one-hot encoding.

One-hot encoding describes the encoding where every $\ket{e}_{\text{L}}$ of the $2^n$ logical $n$-qubit basis states with
\begin{equation}
    \ket{e}_{\text{L}} \in \left\{ \bigotimes_{i=1}^{n}\ket{b_i}_{\text{L}} \:\bigg|\:  b_i \in \{0, 1\}, i \in [1,n], n \in \mathbb{N} \right\},
\end{equation}
is encoded in $2^n$ physical states.
To this end, all physical states are zero except the one corresponding to the index $j \in [0, 2^n-1]$ of the given basis state $\ket{e}_{\text{L}}$ in a list of all logical basis states.\footnote{The ordering of the list is in principle arbitrary, as long as it is kept consistently.}

Formally, this can be described as
\begin{equation}\label{eq:QuantumOne-hotEncoding}
    \ket{e}_{\text{L}} \equiv \ket{0}^{\otimes j} \otimes \ket{1} \otimes \ket{0}^{\otimes 2^n -1-j}.
\end{equation}

Here, states without subscript $\text{L}$ refer to physical states (\eg Fock states $\ket{0}, \ket{1} \in \mathcal{F}$) rather than logical ones \cite{QuantumMechanicalDescriptionLinearOptics}.

As it turns out, only for one photon as input to a linear optical multiport interferometer on a given number of modes $\neq 0$ all quantum transformations can be implemented with linear optics only. In that case (\ie one-hot encoding with a single photon) the classical scattering matrix of the multiport interferometer and the quantum transformation by the multiport interferometer coincide. Consequentially, the classical scattering experiment performed with a classical one-hot encoding -- \ie a laser instead of a single photon source -- can be used as a stand-in for the real quantum experiment as the resulting scattering matrix will coincide. That means, the quantum stand-in experiment determines the implemented scattering matrix. This is only feasible for a low amount of qubits, because the number of needed modes to use this encoding scales exponentially \cite{LinearOpticsOnlyAllowEveryQuantumOperationForOnePhotonOrOnePath, MethodToDetermineQuantumOperationsRalizableWithLinearOptics}.

Specifically for the quantum equivalent scattering experiment, the classical one-hot encoding can be described mathematically as follows:
\begin{equation}\label{eq:ClassicalOne-hotEncoding}
    \ket{e}_{\text{L}} \equiv \ket{0}^{\otimes j} \otimes \ket{\alpha} \otimes \ket{0}^{\otimes 2^n -1-j},
\end{equation}
where $\ket{0}$ and $\ket{\alpha}$ are coherent states and $j \in [0, 2^n-1]$ as in Eq.~\ref{eq:QuantumOne-hotEncoding}. The parameter 
$\alpha$ is an arbitrary nonzero value proportional to the light intensity, its numeric value is irrelevant, as the intensity distribution is normalized in the analysis.

\subsection{Light Source}

The light source used for the quantum-analogous experiment was a tunable laser, configured to an output wavelength of $\sim\qty{1552,701}{\nano\meter}$ 
(\textit{ITU Channel} 32 $+ \qty{28}{\giga\hertz}$). The claimed optical output power of the laser was in the range of 
$\sim \qtyrange{22,336}{22,387}{\milli\watt}$.
The laser emits a coherent state, which encodes the logical $\ket{1}_{\text{L}}$ on a given input mode. In contrast, the absence of input light on a mode (\ie the vacuum state) encodes the logical $\ket{0}_{\text{L}}$. 
The light was delivered from the direct laser output to the photonic processor via a \qty{5}{\meter} single-mode optical fiber. This fiber passed through a fiber polarization controller, which adjusts the polarization state of the light via stress-induced birefringence in the fiber. The fiber was then connected to the first input mode of the photonic processor.
To maximize coupling efficiency, the polarization controller was tuned to the maximum of power at the detector for the first output mode (see Sec.~\ref{sec:measurementsetup}), while the photonic processor was set to the identity matrix. This ensured optimal light coupling to the processor, as the processor is polarization-sensitive -- \ie it attenuates all power of the misaligned polarization at the inputs which just increases loss and thus would decrease the signal-to-noise-ratio (see Fig.~\ref{UV_Fig}).

\subsection{Measurement Setup}
\label{sec:measurementsetup}

The outputs from all eight modes of the photonic processors were routed as follows: each output port was connected to a \qty{1}{\meter} single-mode fiber, in turn coupled to a single-mode, fiber-coupled photodetector with an \ch{InGaAs} photodiode. These support wavelengths from \qtyrange{1250}{1650}{\nano\meter} and the detectors feature a \qty{20}{\giga\hertz} bandwidth. 

The photodetectors are connected to a multichannel \qty{32}{\bit} \gls{adc} at their \qty{2,92}{\milli\meter} analog output ports via coaxial cables. The \gls{adc} was interfaced with a Raspberry Pi 4B equipped with \qty{2}{\giga\byte} RAM for data acquisition and processing. All photodetectors were also electrically grounded to reduce electrical noise.

\subsection{Data Processing}
\label{sec:dataprocessing}

Before conducting the experiment, the noisefloor of the measurement setup was captured and averaged over around $380$ samples. 

Since the measurement setup employed here captured only an intensity distribution and the experiment features a classical light source, the data must be interpreted slightly differently than in an equivalent quantum experiment using single-photon sources and single-photon detectors. In the classical-analogue experiment conducted, rather than measuring a single sample of the underlying photon distribution every shot, we directly measure the intensity distribution which coincides with the quantum case for this encoding.
The average noisefloor -- \ie a constant offset in the recorded raw voltage values -- was subtracted from the captured raw voltage data. This yields a noise-corrected intensity distribution. This intensity distribution is then normalized. The square root of the normalized intensity for a given mode corresponds to the probability amplitude of the basis state corresponding to this mode in the quantum mechanical sense up to the sign. 

Consequently, the normalized intensity corresponds to the probability (as the probability is the absolute-square of the probability amplitude) and is used in further calculations.

\subsection{Experiment}

For each iteration of the experiment, a general unitary matrix is generated to represent the whole pre-measurement quantum circuit shown in Fig.\,\ref{architecture}. This unitary consists most importantly of a controlled Haar-random unitary matrix (\(U\)) followed by the controlled-not \gls{vqc} unitary (\(U_{\text{VQC}}\)).

\textbf{Assumption:} The unknown unitary \(U\) is assumed to be known for the purpose of constructing a composite unitary that simulates the full quantum channel within a single 8-mode optical processor, however, its internal structure remains inaccessible during the process tomography procedure.

The whole quantum circuit is then implemented as a single unitary on the photonic processor, measured (see Sec.~\ref{sec:measurementsetup}), and the resulting data is processed (see Sec.~\ref{sec:dataprocessing}). 

The input state of the photonic processor, as illustrated in Fig.~\ref{UV_Fig}, corresponds to $\ket{000}_{\text{L}} \equiv \ket{1000\:0000}$ in the analogous quantum experiment. Since the single-output laser has an optimized coupling to the first input mode of the photonic processor it would be disadvantageous to recouple with different input modes for every new input state. Accordingly, to circumvent the recalibration of the input for every measurement the input state was fixed and instead, the input variation was absorbed into the implemented matrix. This way just one physical configuration was necessary.
This also makes it possible to automate the measurement procedure without an optical switch that would introduce additional losses and degrade the signal-to-noise ratio.
As a result, instead of implementing $2^t$  distinct physical input configurations for each unitary,\footnote{$t=2$ in this implementation, since only the inputs $\{\ket{000}_{\text{L}}, \ket{010}_{\text{L}}, \ket{100}_{\text{L}}, \ket{110}_{\text{L}} \}$ are implemented, the control qubit always being $0$, as seen in Fig.~\ref{architecture}.} the input was fixed to $\ket{000}_{\text{L}}$. For each update of the parameters $\{\theta_i, \phi_i \}$, we applied $2^t$ corresponding updated unitaries representing the different input permutations. 
The resulting output data was processed to extract the probabilities, which were then passed to the optimizer as the entries of the implemented matrix. The optimization provides then the next iteration to be implemented. The procedure was iteratively repeated for a predefined number of steps.

\begin{figure*}[t]
\vspace{10mm}
\includegraphics[width=1\textwidth,clip]{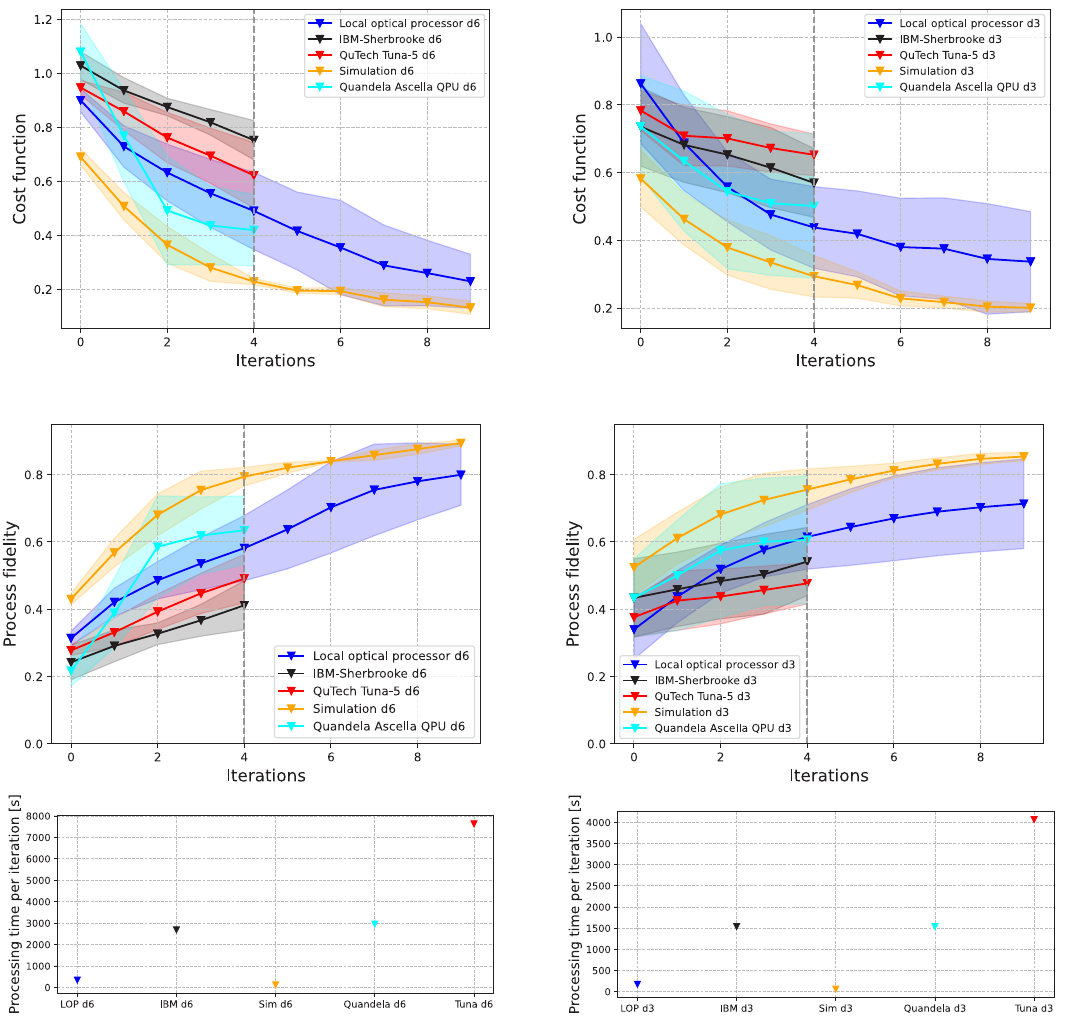}
\caption{Benchmarking results comparing the local optical processor to alternative computational platforms in terms of the cost function (Eq.\,\ref{costfun}), process fidelity (Eq.\,\ref{processfid}), and processing time per iteration for circuit depths $d=3$ and $d=6$. Platforms include a classical simulation, the 6-mode quantum optical processor \textit{Ascella} (Quandela), the 127-qubit superconducting processor \textit{IBM-Sherbrooke}, and the 5-qubit superconducting processor \textit{Tuna-5} (QuTech). All methods were evaluated using the same Haar-random unitaries and initial conditions. Uncertainties in the cost function and process fidelity were obtained from three distinct Haar unitaries, while processing time statistics were computed over five independent iterations. Only the first five iterations are presented for the non-local quantum processors, as access and execution times on these platforms are significantly limited.}
\label{img:result}     
\end{figure*}

\section{Results}
\label{Results}

In Fig.\,\ref{img:result}, we present the performance of the local optical processor described in Fig.\,\ref{UV_Fig} for VQC depths \( d = 3 \) and \( d = 6 \). This performance is benchmarked against several quantum hardware platforms, including the 5-qubit QuTech Tuna, IBM Sherbrooke, and Quandela Ascella processors, as well as a classical simulation for 8194 shots per measurement. Metrics for comparison include process fidelity, cost function values, and processing times.

Uncertainties in the cost function and process fidelity were estimated using three different Haar-random unitaries, while uncertainties in the processing time per iteration were evaluated using five distinct measurements per device.

For the simulation and the local optical processor, results are shown for 10 optimization iterations. In contrast, only 5 iterations were performed on the quantum hardware due to time constraints.

At \( d = 6 \), we observe a steeper improvement in both cost function and process fidelity compared to \( d = 3 \), with the local optical processor achieving fidelities of up to 0.8 after 9 iterations. Notably, the optical quantum/classical processors outperform the superconducting ones in the VQC optimization task. Across both depths, there is a consistent fidelity gap between \qty{0.10}{} and \qty{0.22}{} in favor of the optical platforms after 5 iterations. Increasing the depth from \( d = 3 \) to \( d = 6 \) did not yield performance improvements for superconducting devices, likely due to increased noise from decoherence and dephasing associated with deeper circuits.

Among the optical platforms, Ascella demonstrated slightly better performance than the local optical processor. With one-hot encoding, the results remained within the uncertainty bounds for \( d = 3 \). In simulation, quantum noise effects were found to be negligible relative to the overall thermal noise.

In terms of processing time per iteration, Tuna-5 exhibited significantly longer times, approximately three times that of the IBM hardware. On average, quantum devices required approximately \qty{1500}{\second} per iteration for \( d = 3 \). The shorter processing times of the local optical processor may be attributed to its direct amplitude measurements, eliminating the need for repeated sampling.

\section{Conclusions}
\label{Conclusions}

In this work, we present the first benchmarking study across multiple quantum platforms for the task of variational quantum process tomography, building upon the methodology introduced in~\cite{Galetsky_2024}. We focus on open-access systems and compare their performance to that of our local optical processor, illustrated in Fig.\,\ref{UV_Fig}. The benchmarking spans from standard simulation to hardware implementations, including the Tuna-5 processor developed by QuTech~\cite{valléssanclemente2025optimizingfrequencypositioningtunable}, IBM’s 127-qubit Sherbrooke processor~\cite{mckay2023benchmarkingquantumprocessorperformance}, and the Ascella photonic quantum processor developed by Quandela~\cite{Maring2024}.

Our analysis evaluates three key metrics for variational quantum circuits (VQCs) at depths \( d = 3 \) and \( d = 6 \): the cost function, process fidelity, and processing time per iteration.

We provide a detailed description of the experimental and optical setup used in our processor, including how one-hot encoding and unitary decomposition enable the implementation of the hybrid VQC algorithm on the optical platform.

Our results show that quantum/classical optical processors consistently outperform superconducting devices across both depths. Specifically, we observe a process fidelity difference ranging from \qty{0.10}{} to \qty{0.22}{} between optical and superconducting platforms after five iterations. Moreover, increasing the circuit depth from \( d = 3 \) to \( d = 6 \) leads to improved performance on the optical processors, achieving fidelities up to \qty{0.8}{} after ten iterations. In contrast, the superconducting devices show no significant performance gains at higher depths, likely due to increased susceptibility to decoherence and dephasing noise stemming from deeper circuits.

Due to the limitations of the 8-mode optical processor in independently representing both the unknown unitary \(U\) and the VQC unitary \(U_{\text{VQC}}\), we implemented a single joint unitary that represents the entire quantum channel shown in Fig.\,\ref{architecture}. This approach required assuming prior knowledge of \(U\) in order to construct the composite unitary, as the primary objective of this work is to evaluate the performance of the \gls{vqc} algorithm on quantum hardware. 

To preserve the security of \(U\) with no prior assumptions, we propose in future research of implementing \(U\) and \(U_{\text{VQC}}\) on two separate and independent optical processors -- one encoding a fixed, unknown Haar-random unitary, and the other dedicated to dynamically updating the variational parameters \(\{\theta_i, \phi_i\}\). 

\section{Appendix}
\label{Sec:Appendix}
\subsection{Calibration data and hardware specifics}
\subsubsection{IBM-Sherbrooke}
\begin{figure*}
\includegraphics[width=0.5\linewidth]{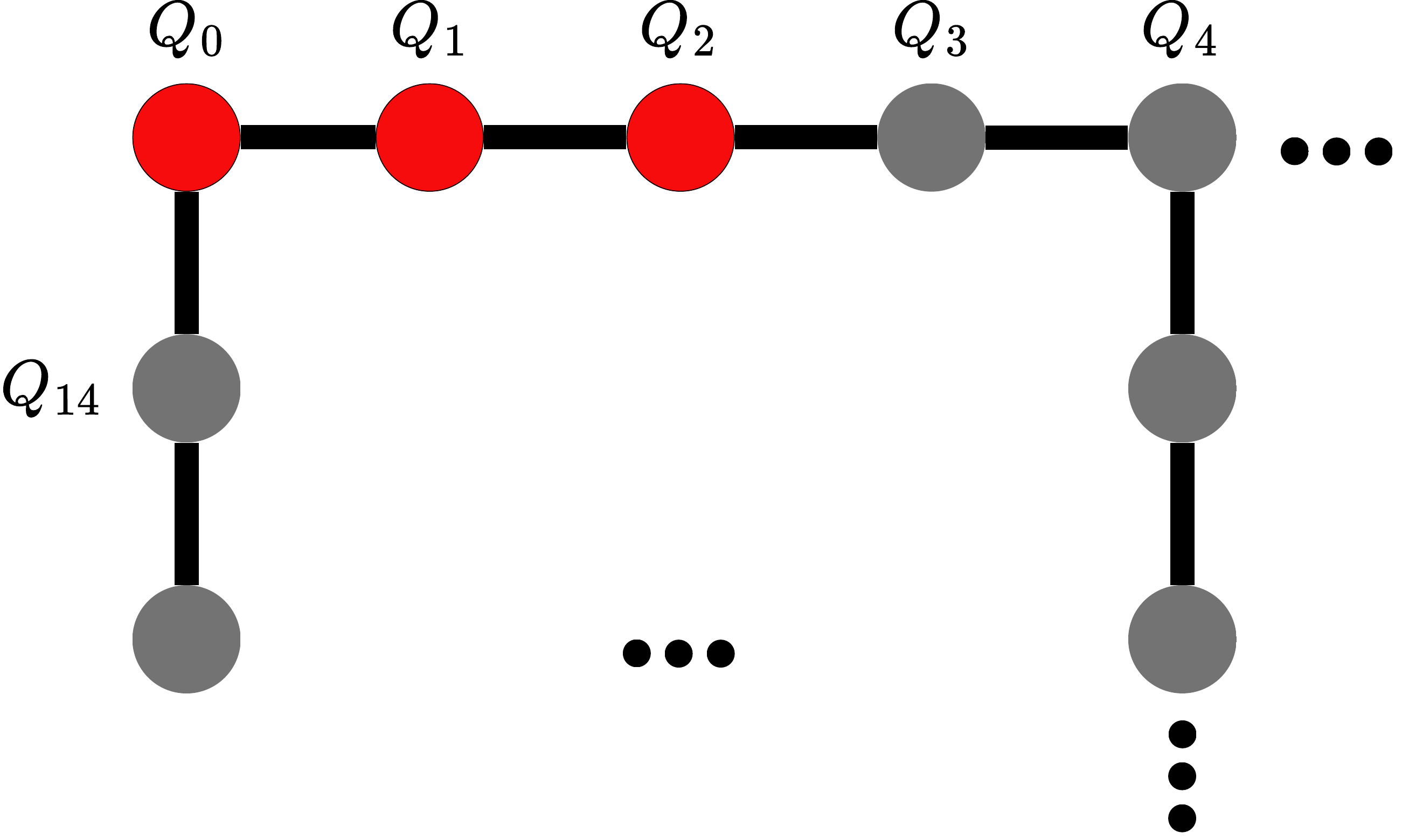}
\caption{Coupling map of IBM Sherbrooke processor and the considered qubits in red for the algorithm.}
\label{fig:IBM}       
\end{figure*}

For IBM superconducting processor, we select the first three qubits from the coupling map as seen in Fig.\,\ref{fig:IBM}. With the calibration data shown in Table~\ref{tab:IBM}. Here $P(1/0\rightarrow0/1)$ is the probability that a qubit prepared in state $1/0$ is measured in the state $0/1$ respectively.  $T_1$ and $T_2$ correspond to the qubit decoherence and dephasing times, respectively. 

\begin{table*}[h!]
\centering
\begin{tabular}{|c|c|c|c|c|c|c|c|c|}
\hline
Qubit & $T_1$ (\qty{}{\micro\second}) & $T_2$ (\qty{}{\micro\second}) & Readout err & P(1$\to$0) & P(0$\to$1) & ID gate err & $\sqrt{x}$ gate err & X gate err \\
\hline
0 & \qty{429.42}{} & \qty{345.20}{} & \qty{0.012}{} & \qty{0.0171}{} & \qty{0.0068}{} & \qty{0.00014}{} & \qty{0.00014}{} & \qty{0.00014}{} \\
1 & \qty{325.44}{} & \qty{227.67}{} & \qty{0.096}{} & \qty{0.0962}{} & \qty{0.0962}{} & \qty{0.00051}{} & \qty{0.00051}{} & \qty{0.00051}{} \\
2 & \qty{236.42}{} & \qty{249.88}{} & \qty{0.137}{} & \qty{0.126 }{} & \qty{0.149 }{} & \qty{0.00019}{} & \qty{0.00019}{} & \qty{0.00019}{} \\
\hline
\end{tabular}
\caption{Calibration data for qubits 0, 1, and 2 from IBM Sherbrooke.}
\label{tab:IBM}
\end{table*}

\subsubsection{Tuna-5 QuTech}

\begin{figure*}
\includegraphics[width=0.20\linewidth]{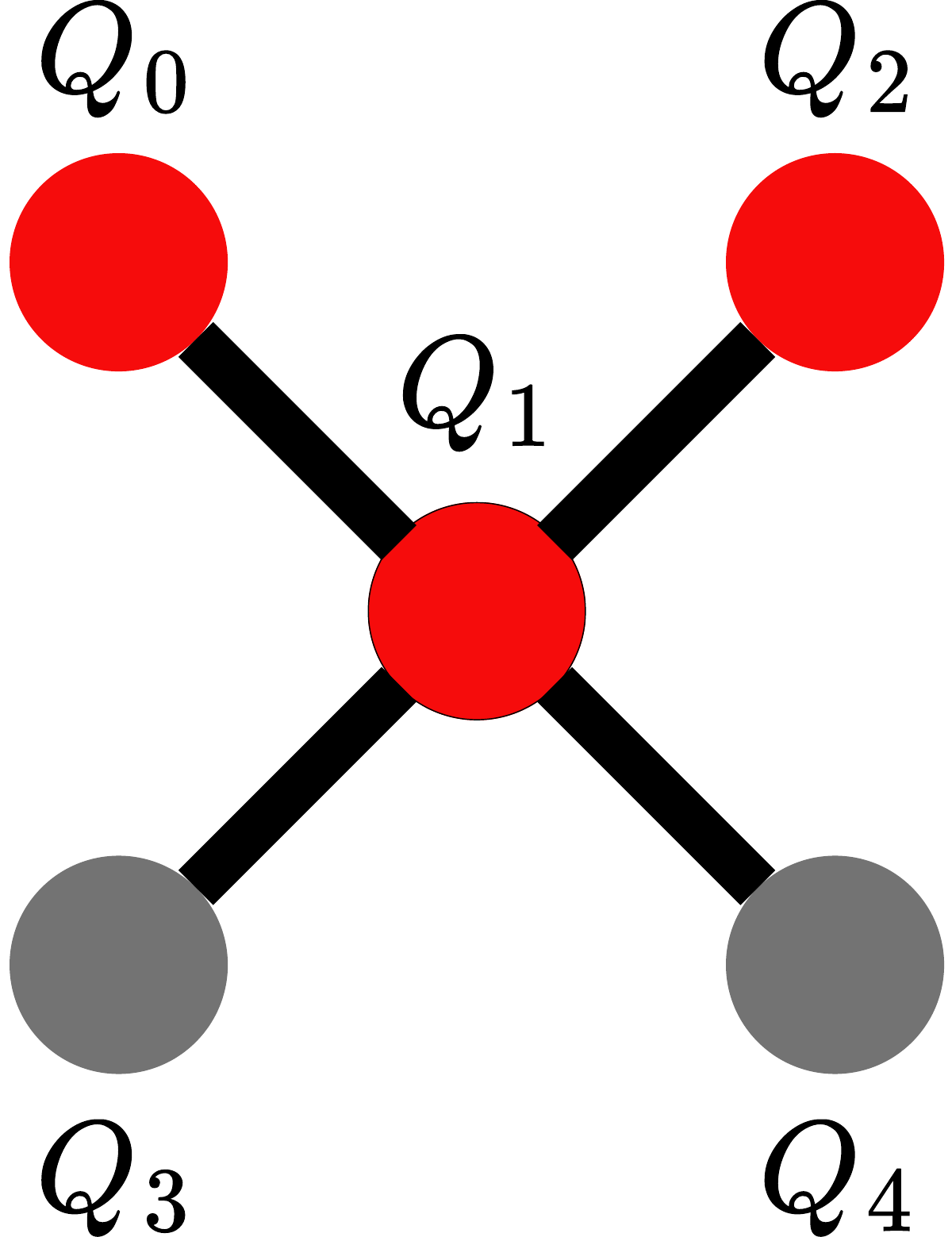}
\caption{Coupling map of Tuna-5 processor and the considered qubits in red for the algorithm.}
\label{fig:Tuna}       
\end{figure*}

Similarly, for the Tuna-5 superconducting processor, we select the first three qubits from its coupling map, as shown in Fig.~\ref{fig:Tuna}. The corresponding calibration data is provided in Table~\ref{tab:Tuna}. Equivalently to the IBM processor, $T_{2R}$ denotes the Ramsey decoherence time.

\begin{table*}[h!]
\centering
\begin{tabular}{|c|c|c|c|c|c|c|c|c|}
\hline
Qubit & $T_1$ (\qty{}{\micro\second}) & $T_{2R}$ (\qty{}{\micro\second}) & Readout err & Init err  & Single gate err & Bell state err \\
\hline
0 & \qty{74.0}{} & \qty{24.1}{} & \qty{0.00870}{} & \qty{0.0000}{} & \qty{0.280}{} & Q[0,2] = \qty{0.015}{} \\
1 & \qty{50.5}{} & \qty{26.0}{} & \qty{0.00350}{} & \qty{0.0013}{} & \qty{0.100}{} & Q[1,2] = \qty{0.012}{} \\
2 & \qty{55.0}{} & \qty{16.4}{} & \qty{0.00350}{} & \qty{0.0016}{} & \qty{0.060}{} & - \\
\hline
\end{tabular}
\caption{Calibration data for qubits 0, 1, and 2 from Tuna-5 processor.}
\label{tab:Tuna}
\end{table*}

\subsubsection{Ascella QPU Quandela}
The Ascella calibration does not directly provide noise parameters beyond those related to state initialization, specifically photon indistinguishability (HOM) and multi-photon emission probability ($g^{2}$). The processor architecture is identical to that of our local system shown in Fig.\,\ref{UV_Fig}, but scaled to support 12 optical modes, of which only the first 6 are utilized in this experiment.

\begin{table*}[h!]
\centering
\begin{tabular}{|l|c|}
\hline
Parameter & Value \\
\hline
Clock Rate (\qty{}{\mega\hertz}) & \qty{80}{} \\
Hong-Ou-Mandel (HOM) Interference (\qty{}{\percent}) & \qty{87.54}{} \\
Transmittance (\qty{}{\percent}) & \qty{1.83}{} \\
$g^{(2)}(0)$ (\qty{}{\percent}) & \qty{1.47}{} \\
Number of Optical Modes & \qty{12}{} \\
\hline
\end{tabular}
\caption{Characterization of the quantum optical Ascella processor.}
\label{tab:quandela}
\end{table*}

\bibliography{bibliography.bib}

\end{document}